# Robust magnetoelectric coupling in altermagnetic-ferroelectric type-III multiferroics

Wei Sun[1], Wenxuan Wang[2]*, Changhong Yang[1]*, Ying Liu[3], Xiaotian Wang[4]*, Shifeng Huang[1], Zhenxiang Cheng[4]*

1 Shandong Provincial Key Laboratory of Preparation and Measurement of Building Materials, University of Jinan, Jinan, 250022, China

2 School of Material Science and Engineering, University of Jinan, Jinan, 250022, Shandong, China

3 Hebei Engineering Laboratory of Photoelectronic Functional Crystals, School of Materials Science and Engineering, Hebei University of Technology, Tianjin 300130, China.

4 Institute for Superconducting & Electronic Materials, Australian Institute of Innovative Materials, University of Wollongong, Innovation Campus, Squires Way, North Wollongong, NSW 2500, Australia

* Corresponding Author

### ABSTRACT

Multiferroic materials, characterized by the coexisting of ferroelectric polarization (breaking spatial inversion symmetry) and magnetism (breaking time-reversal symmetry), with strong magnetoelectric coupling, are highly sought after for advanced technological applications. Novel altermagnets, distinct from conventional magnets, have recently been revealed to exhibit unique spin polarization protected by crystal symmetry, which naturally overcomes the isolation of magnetism from ferroelectrics associated with spatial symmetry. In this study, we propose a novel class of type-III multiferroics, where ferroelectricity and altermagnetism are inherently interlocked by crystal symmetry, setting them apart from conventional multiferroics. Through first-principles calculations, ferroelectric switching is shown to fully invert the spin polarization of altermagnets, equivalent to a 180° reversal of magnetic spin. This strong magnetoelectric coupling is further supported by the magneto-optical Kerr effect, revealing a new class of multiferroics with robust, symmetry-driven magnetoelectric coupling and providing a theoretical foundation for the design of next-generation spintronic devices leveraging altermagnetism.

**INTRODUCTION**

Multiferroics, materials exhibiting both magnetic and ferroelectric order,[1] hold significant potential for information processing and memory applications due to their inherent magnetoelectric coupling.[2,3] However, coupling magnetic and ferroelectric orders remains a major challenge because magnetism arises from broken time-reversal symmetry ($\mathcal{T}$), while ferroelectricity stems from broken space-inversion symmetry ($\mathcal{P}$). This fundamental dichotomy limits direct interactions between the two properties.[4] Traditional approaches to overcome this challenge have relied on type-I and type-II multiferroics. In type-I multiferroics, magnetism and ferroelectricity originate from independent mechanisms, leading to weak magnetoelectric coupling that arises primarily from secondary effects such as magnetic anisotropy or Dzyaloshinskii-Moriya interactions.[5-11] This coupling is often too weak to be effectively controlled by external electric or magnetic fields. In contrast, type-II multiferroics derive ferroelectricity directly from specific magnetic orders, resulting in stronger coupling.[12-15] However, the ferroelectric polarization in these materials is typically small and occurs at temperature far below room temperature, severely limiting practical utility.

Altermagnets, a recently identified class of collinear magnets,[16-23] present a transformative opportunity in this field by challenging the conventional understanding of magnetoelectric coupling based on the coupling between polarization and magnetization. Unlike conventional ferromagnets and antiferromagnets, altermagnets exhibit spin polarization without net magnetization. This unique combination enables advantages such as zero stray fields, making them particularly suitable for spintronic applications, including devices leveraging the spin Hall effect and magneto-optical Kerr effect.[24-28] Altermagnetic behavior arises from the interplay of two opposing spin sublattices connected through rotation ($\mathcal{C}$) or mirror ($\mathcal{M}$) symmetry operations, rather than translational ($\tau$) or inversion ($\mathcal{P}$) symmetries. This symmetry-driven spin polarization alternates across reciprocal space while maintaining zero net magnetization across the Brillouin zone.[17] Recent studies have shown that altermagnetism can be induced by breaking $\mathcal{P}$ symmetry in antiferromagnetic materials through methods such as external electric fields,[29] substrate-induced effects,[30-32] or chemical substitution [33,34]. The mirror or rotational symmetries required for altermagnetism are inherently compatible with the broken $\mathcal{P}$ symmetry of ferroelectricity, opening the possibility for direct coupling between altermagnetism and ferroelectricity.

Notably, ferroelectric switching, driven by partial ion displacement under an electric field, often involves symmetry operations beyond $\mathcal{P}$-only transformations. If this switching is accompanied by mirror or rotational operations associated with altermagnetism, direct manipulation of altermagnetism via ferroelectricity becomes feasible. Specifically, altermagnetism cannot be described solely by conventional magnetic groups but is based on the nonrelativistic spin group formalism $[\mathcal{R}_s||\mathcal{R}_l]$, which produces $E(s, \mathbf{k}) = E(-s, \mathcal{R}_l\mathbf{k})$[16], where the transformation on the left (right) of the double vertical bar acts solely in spin space (spatial space). Here s and $\mathbf{k}$ denote the spin and momentum, and $E(s, \mathbf{k})$ refers to spin- and momentum-dependent bands. Thus, if the polarisation-only switching of the ferroelectric occurs through combined $\mathcal{P}$ and $\mathcal{R}_l^{-1}$ operations, the $[\mathcal{R}_s||\mathcal{R}_l]$ symmetry of altermagnetism can break the isolation of magnetism from ferroelectricity via $\mathcal{P}\mathcal{R}_l^{-1}E(s, \mathbf{k}) = E(s, -\mathcal{R}_l^{-1}\mathbf{k}) = E(-s, -\mathbf{k}) = \mathcal{T}E(s, \mathbf{k})$.

This symmetry-driven mechanism defines a new category of multiferroics — type-III multiferroics — that combine the advantages of type-I and type-II systems. Ferroelectricity and magnetism can occur independently, ensuring functional flexibility, while symmetry-mediated coupling between ferroelectricity and altermagnetism offers robust magnetoelectric interactions. Additionally, the perfectly compensated antiparallel magnetisation of altermagnets provides resilience against external magnetic disturbances, enhancing the stability and performance of spintronic devices.

Our DFT calculations confirm that ferroelectric switching can fully reverse the spin polarization in a bilayer $MnPSe_3$, equivalent to a 180° magnetic spin reversal. This strong magnetoelectric coupling is further corroborated by calculations of the magneto-optical Kerr effect. Together, these findings establish a theoretical framework for the design of next-generation memory devices based on altermagnetism, offering a robust platform for advanced spintronic applications.

**RESULTS**

We constructed a multiferroics by stacking two $MnPSe_3$ layers, as shown in Fig. 1a. In this case, the out-of-plane polarization originates from the sliding ferroelectricity by the stacking of layers of van der Waals crystals,[35-39] which can be switched by the lateral sliding of one of the layers. Since the "sliding ferroelectricity" can be constructed without additional intrinsic ferroelectric components, it is an ideal model for constructing multiferroic. In addition, the Mn ions in the MnPSe$_3$ layer contribute a magnetic moment of about 5 μB, which behaves as an easy-plane Néel-type antiferromagnet, as shown in Fig. S1.

**Altermagnetism generated by stacking antiferromagnetic bilayer**

To generate altermagnetism, two conditions must be met:[16] 1) the system must exhibit zero net magnetization, and 2) the sublattices with opposite spins should be connected via rotational or mirror symmetry. For a MnPSe$_3$ monolayer with space group $P\bar{3}1m$, a simplified model of Néel-type antiferromagnetic ordering is shown in Fig. 1b. In this configuration, opposite spin sublattices are connected by inversion or mirror symmetry, denoted as $[\mathcal{C}_2||\mathcal{P}]$ or $[\mathcal{C}_2||\mathcal{M}]$. Additionally, collinear magnets exhibit spin-only symmetry $[\bar{\mathcal{C}}_2||\mathcal{T}]$, leading to $[\bar{\mathcal{C}}_2||\mathcal{T}]$E(s, **k**) = E(s, -**k**),[16,21,40]. This implies that E(s, **k**) is invariant under $\mathcal{P}$ and $\tau$ operations, which include $\mathcal{C}_{2z}$ and $\mathcal{M}_z$ operations equivalent to $\mathcal{P}$ and $\tau$ in 2D systems. When combined with the $[\mathcal{C}_2||\mathcal{P}]$ symmetry described above, the symmetry $[\mathcal{C}_2||\mathcal{P}][\bar{\mathcal{C}}_2||\mathcal{T}]$ enforces E(s, **k**) = E(-s, **k**), resulting in fully spin-compensated bands. This symmetry is equivalent to the combined $\mathcal{PT}$ operation, which preclude the possibility of altermagnetism in monolayer with Néel-type magnetic configuration.[41]

However, the situation changes dramatically in the bilayer system, where the introduction of a layer degree of freedom alters the symmetry, as shown in Fig. 1b. In this case, inversion symmetry connects identical spins rather than opposite spins, breaking the $[\mathcal{C}_2||\mathcal{P}]$ symmetry, but the $[\mathcal{C}_2||\mathcal{M}]$ symmetry along the *c*-axis remains unaffected by the layer stacking. This $[\mathcal{C}_2||\mathcal{M}]$ symmetry connects sublattices with opposite spins, producing E(s, **k**) = $[\mathcal{C}_2||\mathcal{M}]$E(s, **k**) = E(-s, $\mathcal{M}$**k**), which plays a crucial role in generating altermagnetism.

Figures 1c and 1d illustrate the spin-resolved three-dimensional band structures of MnPSe$_3$ monolayers and bilayers. In the monolayers, the spin-up and spin-down bands are degenerate, while in the bilayer system, these bands are splitting. This result validates the symmetry analysis above and demonstrates the feasibility of inducing altermagnetism through layer stacking. Notably, since the generation of altermagnetism is entirely dependent on crystal symmetry, this presents a general mechanism for producing altermagnetism.

**Sliding ferroelectricity of bilayer MnPSe₃**

To confirm the relationship between stacking configurations and polarization, we present three high-symmetry structures for bilayer MnPSe₃ stacking in Fig. 2a and Fig. S2. The stacking configuration can be switched cyclically by relative lateral shift along the vector $r = \frac{1}{3}(a + b)$. In AA' stacking configuration, the lower layer is fully eclipsed by the upper layer, causing pairs of bulky Se atoms to align directly on top of each other, which leads to increased steric repulsion.[42] DFT calculations show that the AA' stacking has an energy 230.45 meV higher than of the AB or BA stacking, making it an unstable configuration. In the AB stacking, the interfacial Se atoms in the lower layer are positioned at the hollow sites of the upper layer, resulting in an out-of-plane polarization of 0.14 pC/m, as shown in Figs. 1a and 2a. Since BA stacking can be obtained by inverting the AB stacking, their polarizations are opposite and can be switched by a relative lateral sliding between the layers, as shown in Fig. 2b. Figure 2c shows that the transition between AB and BA stacking requires overcoming an energy barrier of 30.6 meV/f.u., demonstrating the potential for high-speed switching with a low electric field. The polarization in bilayer MnPSe₃ is generated by the broken $\mathcal{P}$ symmetry due to the layer stacking. This occurs because the Se atoms at the interface between the two layers experience different environments, leading to uncompensated interlayer vertical charge transfer and the formation of an out-of-plane dipole, as depicted in Fig. 2d.

**Altermagnetic-ferroelectric magnetoelectric coupling**

We have shown that layer stacking induces both altermagnetism and sliding ferroelectricity. In the following, we discuss the effect of altermagnetic-ferroelectric magnetoelectric coupling in bilayer MnPSe₃. To describe the system response, we introduce two order parameters, **L** and **P**, where **L** represents the magnetic ordering and **P** represents the out-of-plane polarization. In this context, the $\mathcal{T}$ operation only reverses the **L** to produce $\mathcal{T}$(**+L, +P**) = (**-L, +P**), which behaves as a 180° inversion of the magnetic spin. Note that for sliding ferroelectrics, the polarization switching achieved by relative sliding from the AB stacking to the BA stacking occurs through the combined $\mathcal{PC}_2\mathcal{M}^{-1}$ operation, rather than just $\mathcal{P}$ operation, as shown in Fig. 2d and discussed in more detail in the Supplementary Information. Therefore, we label the AB and BA stacking as (**+L, +P**) and (+L, -P), respectively, and the polarization of (+L, +P) can switch to (+L, -P) via the $\mathcal{P}\mathcal{C}_{2}\mathcal{M}^{-1}$ operation. Moreover, the spin splitting sign of the altermagnets is determined by the spin group rather than the antiferromagnetic order, which is defined as **S**.

Figure 3a shows the spin-dependent differential charge density and Fermi surface of the (+L, +P) state, which clearly demonstrates that the introduction of the ferroelectric dipole causes a directional transfer of spin electrons along the *c*-axis, leading to further manipulation of the altermagnetism, as shown in Fig. 3d. In general, the reversal of **S** can be obtained by reversing **L**. Since the direction of spin splitting defined by **S** results from $\mathcal{T}$ symmetry breaking,[17] it can be reversal by the $\mathcal{T}$ operation, producing $\mathcal{T}$**S** = -**S**, which corresponds to $\mathcal{T}$(+L, +P) = (-L, +P), as shown in Fig. 3b and e.

When the ferroelectric polarization is switched to (+L, -P) by relative lateral sliding between the layers, the inversion of the polarization triggers a reverse spin-electron transfer, as shown in Fig. 3c. Surprisingly, the spin splitting direction in reciprocal-space of the (+L, -P) is the same as that of (-L, +P), despite the absence of symmetry operations on the spin, as shown in Fig. 3e. This is equivalent to a 180° reversal of the magnetic spin. As mentioned earlier, **P** is reversed by the $\mathcal{P}\mathcal{C}_{2z}\mathcal{M}^{-1}$ operation, and E(s, **k**) in the 2D system is invariant under the $\mathcal{C}_{2z}$ operation, thus:

$$\mathcal{P}\mathcal{M}^{-1}E(s, \mathbf{k}) = E(s, -\mathcal{M}^{-1}\mathbf{k}) \tag{1}$$

Combining the $[\mathcal{C}_2\|\mathcal{M}]$ symmetry of altermagnetism,

$$E(s, -\mathcal{M}^{-1}\mathbf{k}) = [\mathcal{C}_2\|\mathcal{M}]E(s, -\mathcal{M}^{-1}\mathbf{k}) = E(-s, -\mathbf{k}) \tag{2}$$

the E(s, **k**) can then be expressed as:

$$\mathcal{P}\mathcal{M}^{-1}E(s, \mathbf{k}) = E(-s, -\mathbf{k}) = \mathcal{T}E(s, \mathbf{k}) \tag{3}$$

This clearly shows that reversing only the ferroelectric vector has the same effect as reversing only the magnetic spin in altermagnets, i.e. $\mathcal{T}$**S** = $\mathcal{P}\mathcal{M}^{-1}$**S** = -**S**. As a result, since **S** depends on both $\mathcal{P}\mathcal{M}^{-1}$ and $\mathcal{T}$ operations, ferroelectricity originating from broken $\mathcal{P}$ symmetry is coupled with spin splitting arising from broken $\mathcal{T}$ symmetry via the $[\mathcal{C}_2\|\mathcal{M}]$ symmetry, even though the altermagnetic and ferroelectric orders occur independently in bilayer MnPSe$_3$.

Therefore, for altermagnets with $[\mathcal{R}_s\|\mathcal{R}_l]$ symmetry, if the switching of the ferroelectric vectors can occurs through the combined $\mathcal{P}\mathcal{R}_l^{-1}$ operation, the altermagnetism can be directly manipulated via: $\mathcal{P}\mathcal{R}_l^{-1}E(s, \mathbf{k}) = E(s, -\mathcal{R}_l^{-1}\mathbf{k}) = E(-s, -\mathbf{k}) = \mathcal{T}E(s, \mathbf{k})$. Additionally, the switching of the ferroelectric vectors can include additional $\mathcal{C}_{2z}$ and $\mathcal{M}_z$ operations that are invariant to E(s, **k**) in the 2D system. We also designed the SnS$_2$/MnPSe$_3$/SnS$_2$ multiferroic heterostructure, whose polarization switching is achieved only by $\mathcal{PM}^{-1}$ operations. This structure exhibits consistent magnetoelectric coupling, further confirming our results, as shown in Fig. S4. This demonstrates an unprecedented method for establishing strong magnetoelectric coupling in type-III multiferroics, distinct from conventional multiferroics.

**Magneto-optical Kerr effect**

We further calculated the magneto-optical Kerr effect of the MnPSe$_3$ bilayer, a widely used sensitive probe for detecting $\mathcal{T}$ symmetry breaking and spin splitting.[43-45] For the MnPSe$_3$, bilayer with easy-plane anisotropy, the general form of the dielectric tensor, as dictated by the symmetry operation of the magnetic space group *Cm*, is expressed as:

$$\sigma = \begin{bmatrix} \sigma_{xx} & 0 & \sigma_{xz} \\ 0 & \sigma_{yy} & 0 \\ \sigma_{zx} & 0 & \sigma_{zz} \end{bmatrix} \qquad (4)$$

This tensor has five independent coefficients. The complex Kerr angle in the film limit is then calculated using the equation:[46,47]

$$\phi_K = \theta_K + i\eta_K = \frac{Z_0 d(\sigma_{xz} - \sigma_{zx})}{1 - (n_s + Z_0 d \sigma_{xx})^2} \qquad (5)$$

Here, the real part $\theta_K$ represents the Kerr rotation angle, while the imaginary part $\eta_K$ corresponds to the Kerr ellipticity. $n_s$ is the refractive index of the substrate, $Z_0$ is the impedance of free space, and *d* the thickness of the MnPSe$_3$ bilayer.

The calculated magneto-optical Kerr effect is shown in Fig. 4a. A pronounced Kerr signal is observed in the (**+L, +P**) state, attributed to the anisotropic optical conductivity introduced by the $\mathcal{T}$ symmetry breaking. Moreover, the Kerr angle reverses with the polarization-dependent spin splitting reversal, as illustrated in Fig. 4c. The spin texture at the highest valence band highlights the correlation between ferroelectric switching and magneto-optical Kerr effect responses, as shown in Fig. 4b and d. Due to strict symmetry constraints, the spin texture can be reversed by ferroelectric switching, resulting in an opposite Kerr signal. The magneto-optical Kerr effect for the (**-L, +P**) state was also calculated, yielding results consistent with the (**+L, -P**) state. This ferroelectric control of the Kerr signal further demonstrates the robustness of the magnetoelectric coupling in ferroelectric-altermagnetic multiferroics.

In summary, stacking bilayer van der Waals crystals induces both altermagnetism and sliding ferroelectricity, resulting in robust magnetoelectric coupling effects governed by lattice symmetry constraints. Our symmetry analysis reveals direct coupling between $\mathcal{P}$ symmetry breaking and $\mathcal{T}$ symmetry breaking via altermagnetism. This interaction establishes a type-III multiferroic system, characterized by the coexisting of ferroelectric polarization and altermagnetism intrinsically linked through symmetry operations, distinct from conventional multiferroics. The magneto-optical Kerr effect further confirms that ferroelectric switching in bilayer MnPSe₃ is equivalent to a 180° reversal of the magnetic spin, showcasing an unprecedented form of magnetoelectric coupling.

**METHODS**

The atomic properties and electronic structure of the materials were calculated using first-principles simulations within density functional theory (DFT)[48,49]. The projected augmented wave pseudopotentials method, as implemented in the Vienna *Ab* initio Simulation Package (VASP)[50,51] was employed. The exchange-correlation energy was calculated using the generalized gradient approximation (GGA) of the Perdew-Burke-Ernzerhof form[52], and the plane wave cutoff energy was set to 500 eV. A Hubbard $U_{eff}$ = 5 eV with the Dudarev parametrization was applied to properly describe the localization of Mn 3$d$ orbitals.[53] The semiempirical DFT-D3 method was used to include the van der Waals interaction[54]. For MnPSe₃ calculations, a centered 9 × 9 × 1 Monkhorst-Pack $k$-point mesh was used.[55] To eliminate periodic boundary effects, the vacuum space between adjacent slabs was set to exceed 15 Å along the $z$ direction. Using the conjugate gradient method, the plane lattice constant and atomic coordinates were fully relaxed until the energy and force converged to $10^{-6}$ and $10^{-2}$ eV/Å, respectively. The Berry-phase method was employed to evaluate polarization magnitude,[56] and the ferroelectric transition switching pathway was obtained using the climbing image nudged elastic band (CI-NEB) method.[57] The dielectric function $\varepsilon_{ij}(\omega)$ was calculated using spin-orbit coupling, and the optical conductivity $\sigma_{ij}(\omega)$ was derived using the relation $\varepsilon_{ij}(\omega) = \delta_{ij} + i\frac{4\pi}{\omega}\sigma_{ij}(\omega)$.[44,58]

**ACKNOWLEDGEMENTS**

This work was supported by the National Natural Science Foundation of China (Grant No. 12304141), the Shandong Provincial Natural Science Foundation (Grant No. ZR2023QA001), the Taishan Scholars Program (Grants No. tsqn202312209 and No. tstp20221130), the Shandong provincial key research and development plan (Grant No. 2022CXPT045).

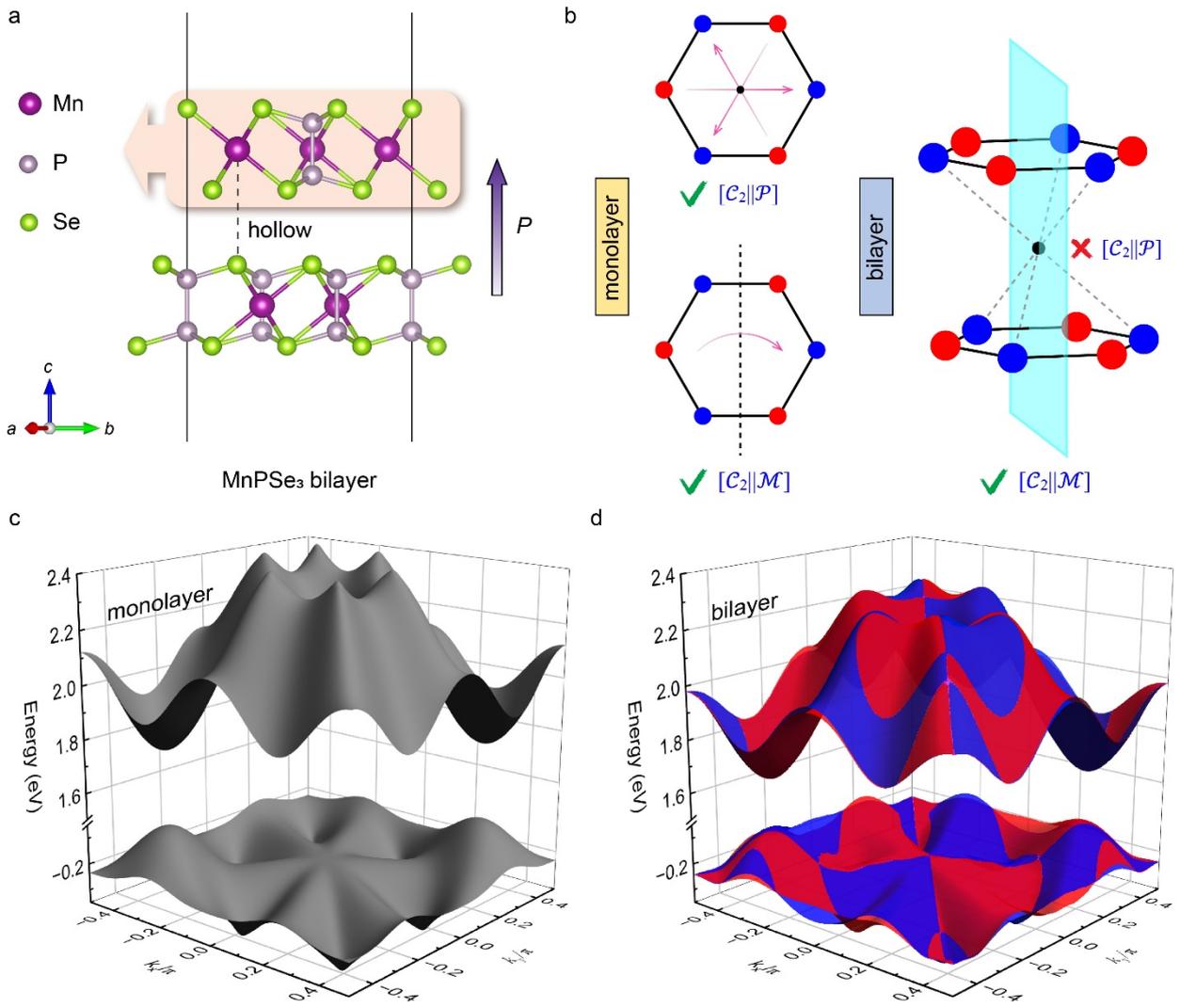

Fig. 1. (a) Side view of MnPSe$_3$ bilayer. (b) Schematic diagrams of Néel-type magnetic ordering for monolayer and bilayer MnPSe$_3$, with (c) and (d) being their corresponding three-dimensional energy band structures, respectively. Red and blue dots (bands) in (b) and (d) indicate the spin-up and spin-down components, respectively, while gray bands in (c) indicate spin degeneracy.

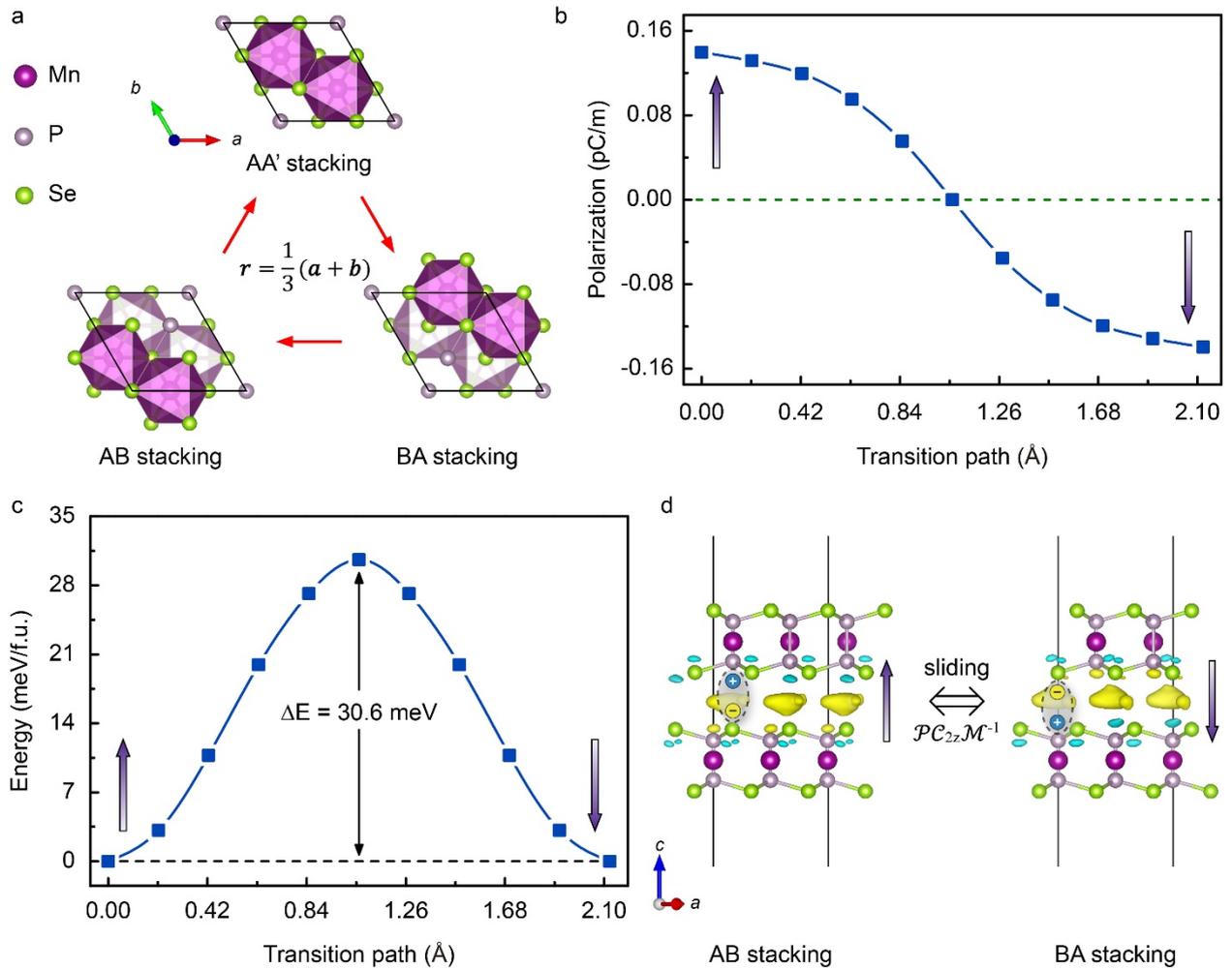

Fig. 2. (a) Top view illustration of MnPSe$_3$ bilayer. The cyclic switching between three stacking configurations can be achieved through the relative displacement vector ***r*** between layers. (b) and (c) are the polarization magnitude and transition energy barriers in the ferroelectric switching path, respectively. The arrow represents the polarization direction. (d) Differential charge density of MnPSe$_3$ bilayer, where the yellow and cyan regions correspond to charge accumulation and depletion.

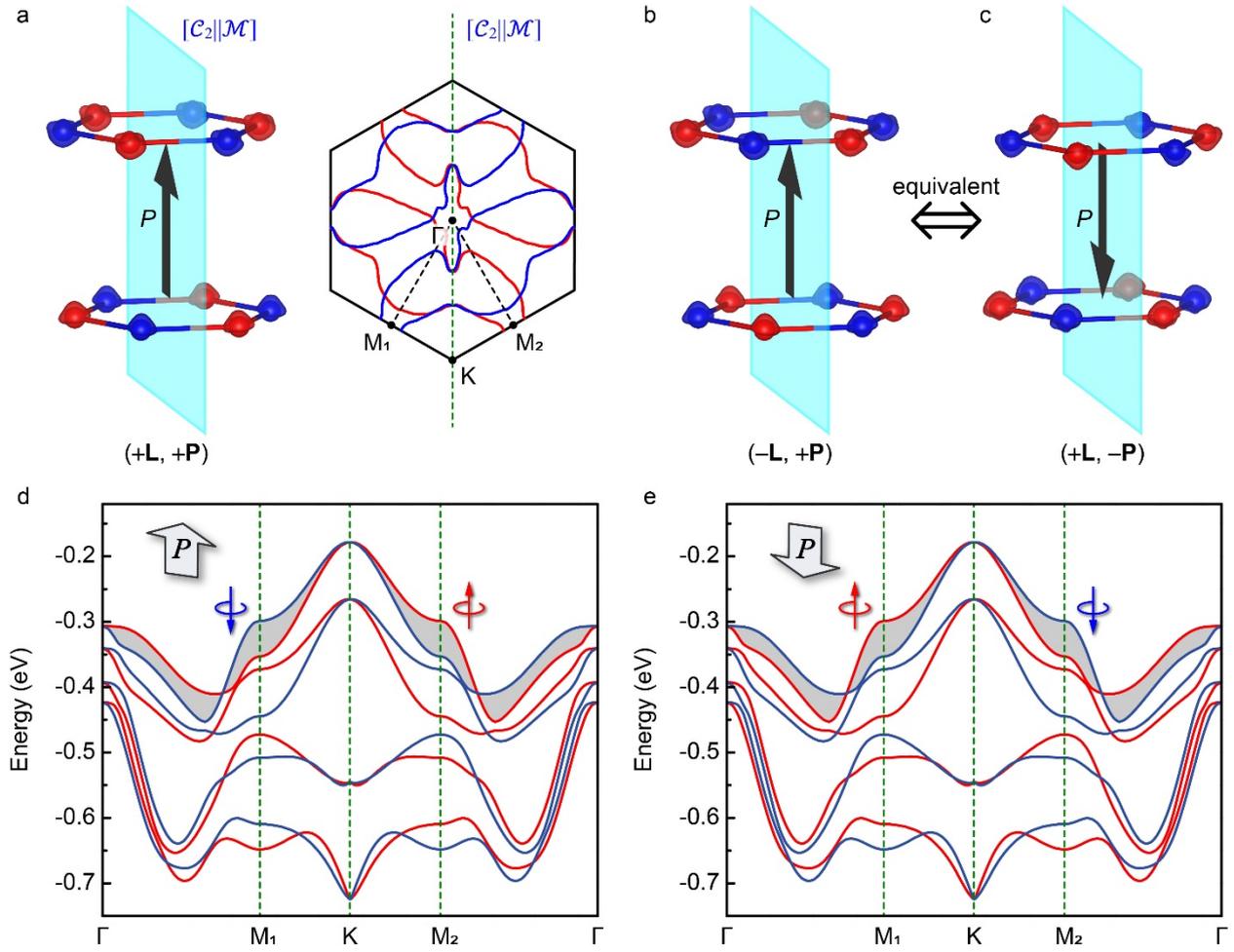

Fig. 3. (a) The spin-dependent differential charge density and Fermi surface at energy = -0.24 eV in first Brillouin zone of (+**L, +P**) state, (c) is its corresponding energy band structure. (b) The spin-dependent differential charge density of (+**L, −P**) and (−**L, +P**), and (d) is their corresponding energy band structure. The black arrows indicate that their energy bands are identical.

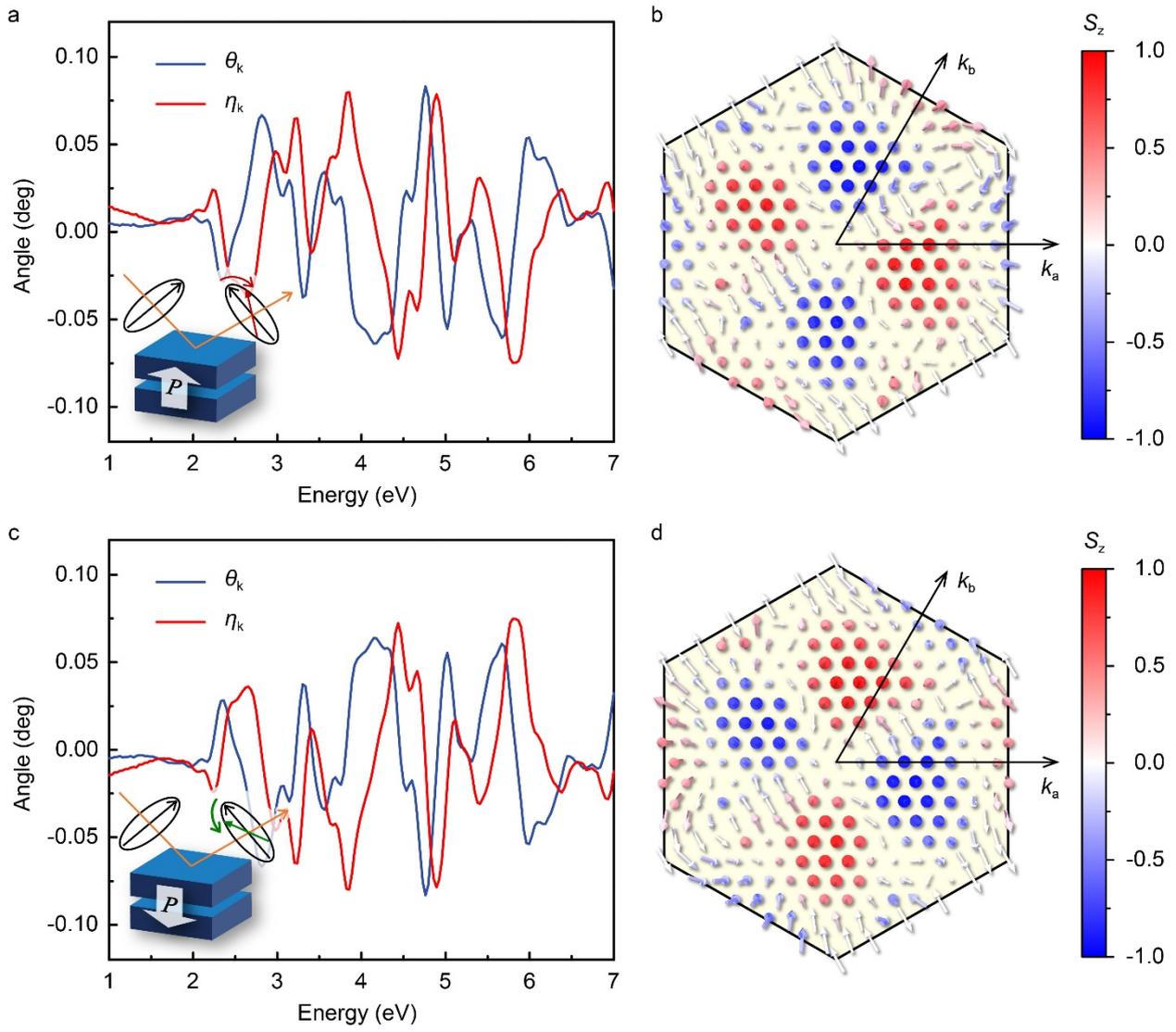

Fig. 4. (a) and (b) are the magneto-optical Kerr signal and spin texture of (**+L, +P**) state, respectively, and the corresponding results for (**+L, −P**) state are (c) and (d).

# Supplementary Information

**Part 1. The band structure and magnetic anisotropy energy of MnPSe$_3$ monolayer.**

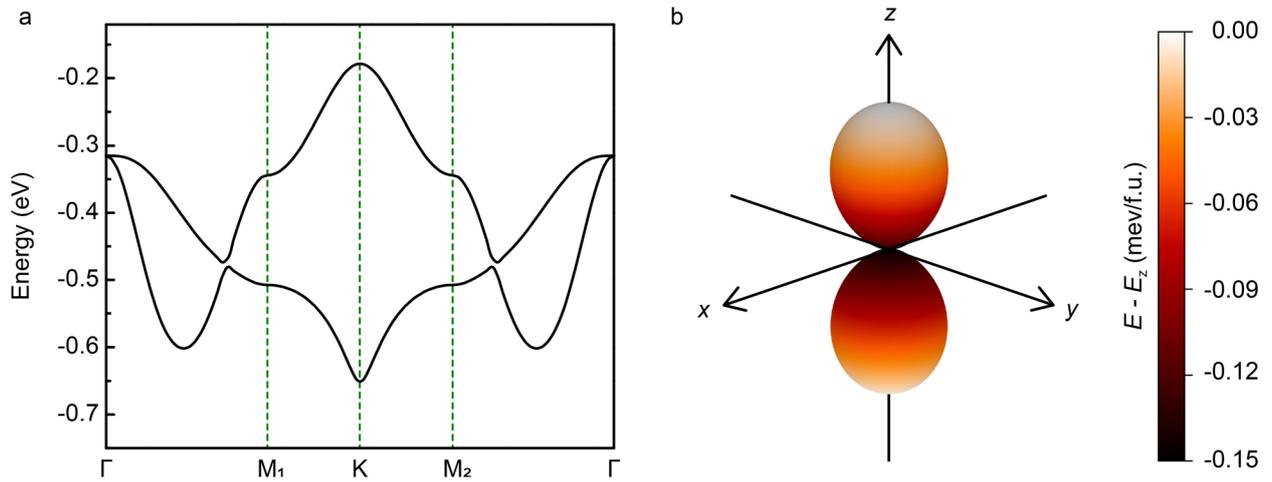

Fig. S1. (a) the band structure of the MnPSe$_3$ monolayer. (b) Angular dependence of the magnetic anisotropy energy of MnPSe$_3$ monolayer with the direction of magnetization lying on the whole space.

**Part 2. Three high-symmetry structures for bilayer MnPSe$_3$ stacking.**

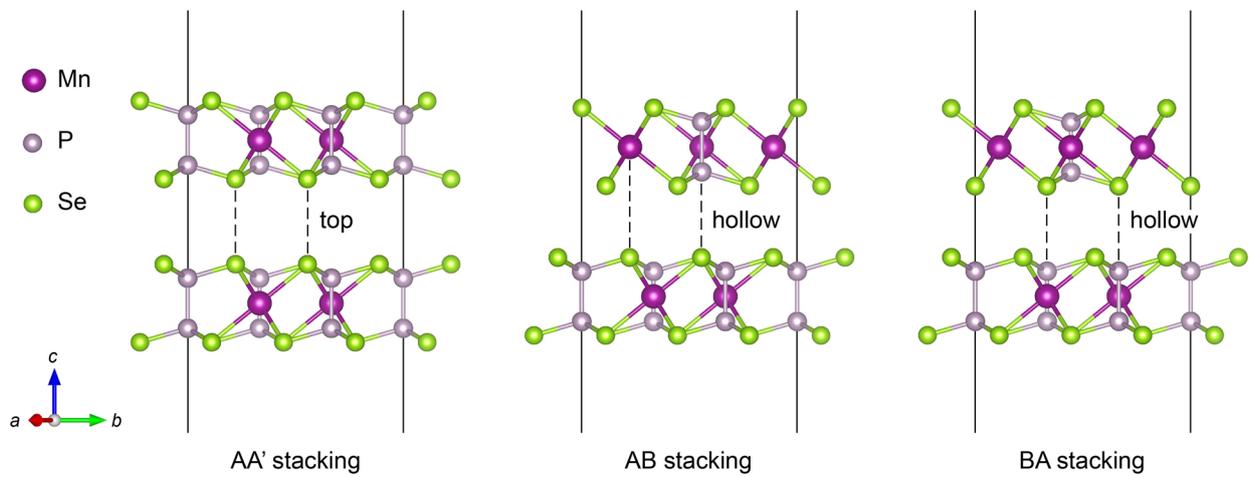

Fig. S2. Side view of three high-symmetry structures for bilayer MnPSe$_3$ stacking.

**Part 3. Switching of sliding ferroelectric polarization in bilayer MnPSe₃ through spatially symmetric operation.**

We first consider a non-magnetic system, where ferroelectric switching realized through interlayer relative sliding is not structurally equivalent to a $\mathcal{P}$-only operation but occurs in combination with a $\mathcal{C}_{2z}$ operation, as shown in Fig. S3a-b. However, in a system that includes magnetic ordering, although the combined $\mathcal{P}\mathcal{C}_{2z}$ operation achieves a reversal of polarization, the magnetic ordering after the symmetric operation is not equivalent to that resulting from actual sliding. To restore the magnetic ordering, the operation must also be combined with the necessary $\mathcal{M}^{-1}$ operation, as shown in Fig. S3c-d.

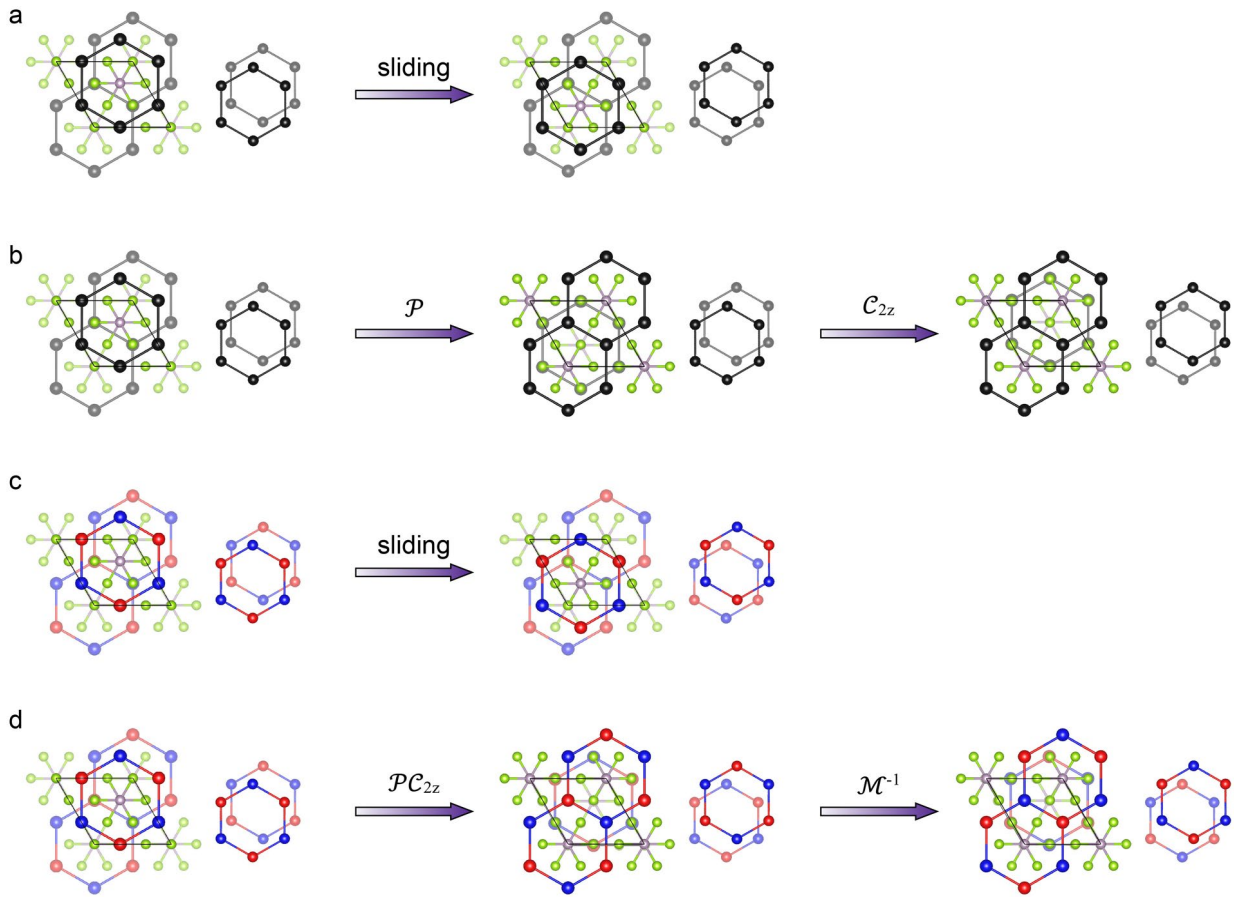

Fig. S3. Ferroelectric switching of bilayer MnPSe₃ by (a) actual slipping and (b) symmetry operates without considering magnetic. (c) and (d) are the corresponding results with magnetic ordering included. The right hexagon highlights the stacking features of the bilayer.

## Part 4. Magnetoelectric coupling effects in SnS$_2$/MnPSe$_3$/SnS$_2$ multiferroic heterostructure.

To further prove our argument, we constructed the SnS$_2$/MnPSe$_3$/SnS$_2$ multiferroic heterostructure, in which polarization switching is achieved by $\mathcal{PM}^{-1}$ without the need for additional symmetry operations, as shown in Figs. S4a and S5. The ferroelectric vector of the SnS$_2$/MnPSe$_3$/SnS$_2$ heterostructure switches though the lateral sliding of the upper and lower SnS$_2$ layers. The polarization transition barrier, calculated using the climbing image nudged elastic band method, is 53 meV, as shown in Fig. S4b. These results demonstrate a robust magnetoelectric coupling between ferroelectricity and altermagnetism via $\mathcal{PM}^1$E(s, **k**) = E(s, -$\mathcal{M}^{-1}$**k**) = E(-s, -**k**) = $\mathcal{T}$E(s, **k**), as shown in Fig. S4c.

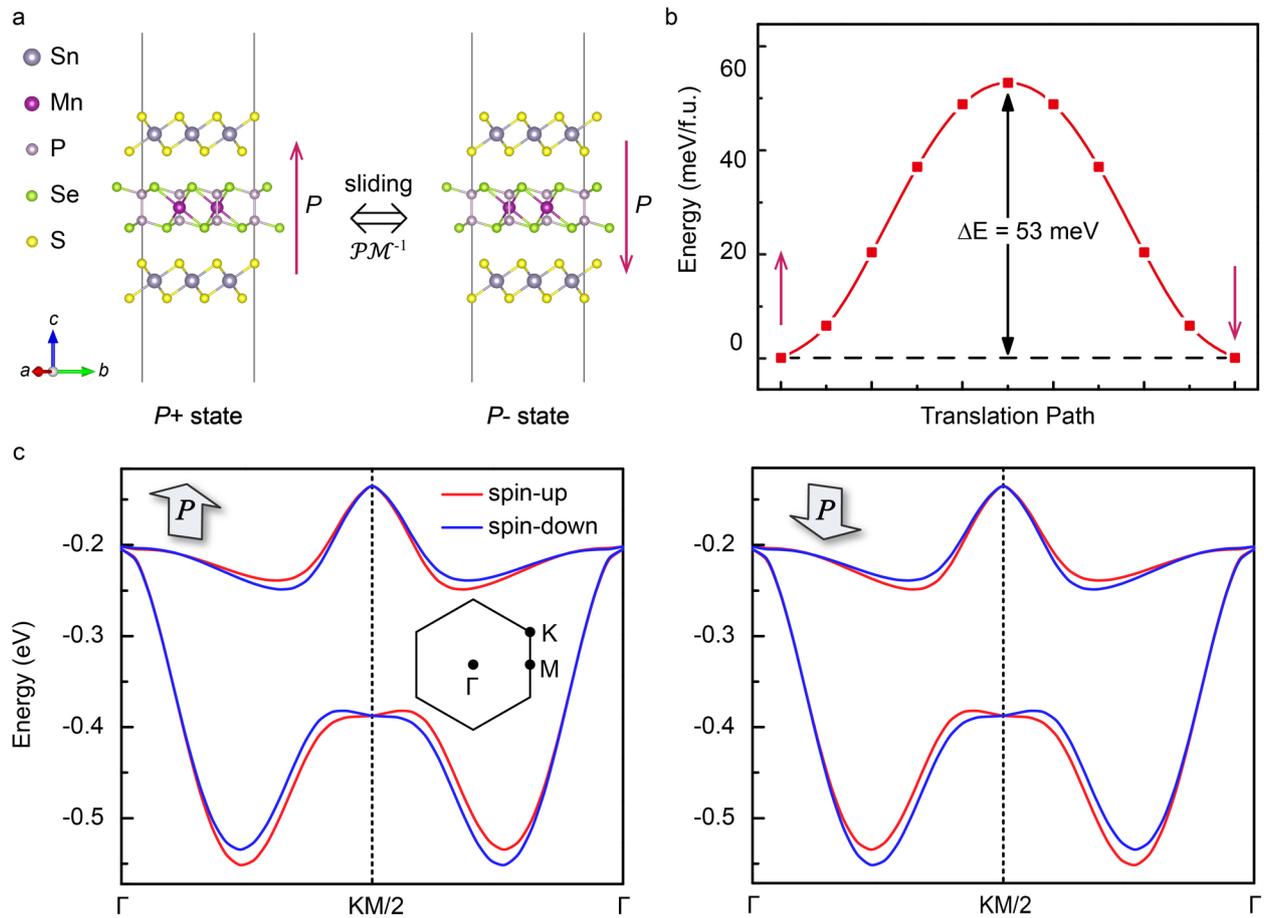

Fig. S4. (a) Side view of the SnS$_2$/MnPSe$_3$/SnS$_2$ heterostructure. (b) Transition energy barriers for ferroelectric switching in the heterostructure. (c) Band structure under opposite polarization state.

Next, we discuss the symmetry operation of the SnS₂/MnPSe₃/SnS₂ heterostructure. In simple non-magnetic structures, polarization reversal achieved by layer sliding is fully equivalent to the $\mathcal{P}$-only operation. However, in the case of Néel-type magnetic ordering, the $\mathcal{P}$ operation simultaneously reverses the magnetic spin. Therefore, the necessary $\mathcal{M}^{-1}$ operation must also be combined to restore the magnetic ordering.

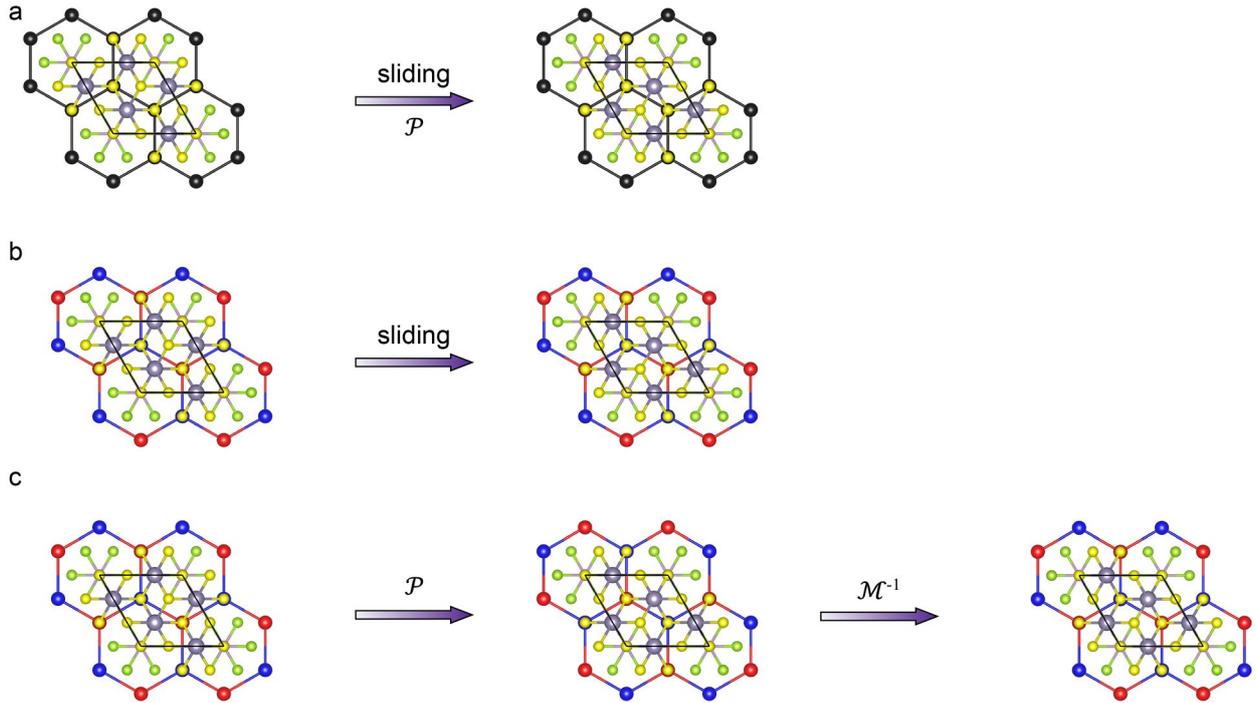

Fig. S5. (a) Ferroelectric switching of SnS₂/MnPSe₃/SnS₂ heterostructure equivalent to $\mathcal{P}$ operation in non-magnetic conditions. (b) and (c) are the ferroelectric switching by actual slipping and symmetry operate respectively in the case of inclusion of magnetic ordering.